\documentclass{article}
\usepackage{amssymb}
\usepackage{amsmath} 
\usepackage{epsfig}
\numberwithin{equation}{section}
\begin{document}

\title{The Zakharov-Shabat spectral problem on the semi-line:
Hilbert formulation and applications}
\author{ J. Leon, A. Spire \\
{\em Physique Math\'ematique et Th\'eorique, CNRS-UMR5825,}\\
 Universit\'e Montpellier 2, 34095 MONTPELLIER (France)}
\date{}\maketitle

\begin{abstract} The  inverse spectral transform for the Zakharov-Shabat
equation on the semi-line $x>0$ is reconsidered as a Hilbert problem. The
boundary data induce an essential singularity as $k\to\infty$ to one of the
basic solutions. Then solving the inverse problem means solving a Hilbert
problem with particular prescribed behavior. It is demonstrated that the direct
and inverse problems are solved in a consistent way as soon as the spectral
transform vanishes as ${\cal O}(1/k)$ at infinity in the whole upper half plane
(where it may possess single poles) and is continuous and bounded on the real
$k$-axis. The method is applied to stimulated Raman scattering and sine-Gordon
(light cone) for which it is demonstrated that time evolution conserves the
properties of the spectral transform.  \end{abstract}

\section{Introduction}

Since the discovery of the inverse spectral transform (IST) in 1967 \cite{ggkm}
to solve nonlinear evolutions (Korteveg-de Vries and later nonlinear
Schr\"odinger \cite{zs}, then sine-Gordon, etc... \cite{akns}) with Cauchy
conditions on the infinite space (and vanishing asymptotic boundary values),
the application of the method to boundary value problems on semi-infinite (or
finite) intervals has been the object of intense research with few general
results, which is reminiscent from the fact that even for linear partial
differential equations, boundary value problem much differ from one another. 
It would be useless to quote all papers that has dealt with boundary value
problems for nonlinear evolutions, only the most recent ones will be cited
here, see \cite{sklyanin}-\cite{hug2} where more complete reference will be
found.  

Taking for instance the nonlinear Schr\"odinger equation (NLS) \cite{zs} for a
field $q(x,t)$, saying that it is solvable by IST on the infinite line actually
means that, given Dirichlet condition $q(x,0)$ on the line
$x\in(-\infty,+\infty)$ and Cauchy conditions $q(\pm\infty,t)=0$ and
$q_x(\pm\infty,t)=0$ for any $t>0$, the method IST allows to construct the
solution $q(x,t)$ at any $t$ for all $x$. Although Dirichlet conditions at
$x=\pm\infty$, namely $q(\pm\infty,t)=0$, constitute with $q(x,0)$ a well-posed
problem, the solution of NLS with IST {\em requires} the adjunction of the
Neuman condition $q_x(\pm\infty,t)=0$. This simply results from the fact that
the Lax pair contains explicitely the derivative $q_x$. The point is that this
supplementary constraint is compatible with the rest of the boundary values and
remains compatible at any time. This {\em stability} is a property of the
spectral transform itself that conserves the class of function (potentials)
chosen at $t=0$. For instance if $q(x,0)$ is in the Schwartz space (the
function and all its derivatives are continuous and vanish faster than any
power of $x$ as $x\to\infty$), then the solution $q(x,t)$ obtained by IST is
also in the Schwartz space which makes the method work.

Such is not anymore the case on finite (or semi-infinite) interval. Indeed,
while the constraint $q_x(\pm\infty,t)=0$ is not restrictive when we already
have $q(\pm\infty,t)=0$, it is not so on a finite domain where imposing the
data of both the function and its derivative produce an over-constrained
problem. Then, although NLS keeps many of its {\em integrability properties} on
the finite interval (with mixed Dirichlet-Neumann conditions on the semi-axis)
\cite{sklyanin} \cite{bibikov}, and  B\"acklund transformation can be used to
determine explicit solvable subclasses \cite{bikbaev}, the general problem is
still unsolved.

A detailed study of the Dirichlet boundary problem (data of $q(x,0)$, $q(0,t)$
and $q(\infty,t)=0$) has been performed in \cite{fok-its}  where it has been
shown that the problem of the determination of $q_x(0,t)$ from $q(0,t)$ is
transformed into the problem of determining some {\em missing spectral datum}. 
This approach allows also to construct classes of solutions but not to solve
the problem in general.

Another point of view is presented in \cite{sabat} where the method is based on
the scattering problem on a line in the $(x,t)$-plane in the
Gelfand-Levitan-Marchenko formulation. This is applied to the Korteveg-de Vries
equation and also provides much insight in the boundary-value problem but shows
that indeed the problem is still open.

An interesting approach has been proposed in \cite{fokas} where both operators
of the Lax pair are treated as joint spectral problems. There the
boundary-value problem becomes a Dirichlet condition on a closed domain, namely
data of $q(x,0)$, $q(0,t)$, $q(\infty,t)=0$ and $q(t,\infty)=0$. Applied to NLS
(defocusing case), this method provides the solution as a system of coupled
integral equations which is then difficult to use. But this approach allows for
particular solutions, see also \cite{boutet}, and for large time asymptotic
behaviors. Moreover it has been succesfuly applied to sine-Gordon in the light
cone (SG) \cite{fokas} for which the Lax pair is compatible with the boundary
values, and then extended to boundary values on arbitrary convex domain
\cite{bea}. The method has also proved being very useful to treat linear
partial differential equations with moving boundaries \cite{fok-bea}.  The
method furnishes the solution under the form of a set of Cauchy-Green integral
equations.

There is another case where the general solution has been obtained by IST: the
stimulated Raman scattering equations (SRS) on the semi-axis \cite{js} (the
{\em sharp-line} version on the finite interval has been treated in
\cite{menyuk} by the method of \cite{fokas}). The boundary-value problem there
is an open-end problem and the stability of direct/inverse spectral
transformation allows indeed to get the general solution on the semi-axis (as
well as the finite interval case) which has been succesfully checked on
numerical simulations \cite{bclp}. The method works also for the discrete
integrable version of SRS \cite{blp} and it is based on the approach of the
{\em dispersion relation} \cite{jl} where the principal Lax operator is taken
as the spectral operator and the auxiliary Lax operator is the tool to define
{\em altogether} the dispersion relation and the time-evolution of the spectral
transform. The advantage of such a method is that, in the physically relevant
case of a medium initially at rest, it allows to get all the information on the
nonlinear Fourier spectrum explicitely from the input radiation fields. An
interesting consequence is the possibility to build literaly the spectral
transform out of measurements and compare it to the theory, which has been done
successfuly on numerical simulations \cite{bclp} of the finite interval, as
well as experimental data \cite{cgl} with infinite line damped model.  The
method has also been extended to treat perturbation of SRS \cite{dok-sch},
confirming the validity of approximations used for comparison with experiments
in \cite{cgl}.

As a general result, it has been shown that the dispersion relation will depend
explicitely not only on the boundary values (as for infinite line \cite{jlsrs})
but also on the spectral transform itself hence producing nonlinear time
evolution of spectral data (of Riccati type) \cite{js}.  Interesting enough,
such an evolution induces the motion of the poles (discrete eigenvalues of the
spectral problem, associated to solitons) and thus results in creation (and
anihilation) of solitons due to input boundaries. The same property has been
found for the karpman-Kaup system \cite{hug1} with a remarkable adequation to
numerical simulations \cite{hug2}.

The spectral problem underlying stimulated Raman scattering and sine-Gordon is
the Zakharov-Shabat system on the semi-axis with prescribed potentials values
in $x=0$. We show here that the solution can be written out of a system
integral equations of Cauchy type where the contour of integration differs from
the one working in the infinite line case. This results from the presence of an
essential sigularity in the asymptotic behaviors of one of the Jost solutions,
a problem that has been underestimated in the litterature so far. 

Then we shall begin with solving the direct and inverse spectral problems in
sections \ref{sec:dir-pro} and \ref{sec:hilb-pro}. In doing this we shall
discover that the reconstructed solutions (Jost solutions) must obey a sequence
of properties, named P1 to P9, that are central to the method. This is then
applied to SRS in section \ref{sec:srs} and to sine-Gordon in section
\ref{sec:sg}. In both cases it is demonstrated that the time evolution of the
spectral transform is such that it conserves the constraints necessary to
ensure the 9 fundamental properties of the Jost solutions.  In order to
illustrate the method for the sine-Gordon, we consider the explicitely solvable
case of a piecewise constant boundary datum which generically stands for the
discretization of an arbitrary boundary value.

\section{The direct problem \label{sec:dir-pro}}

\subsection{Basic solutions}

The time $t$ being an external parameter it is omitted in the whole section. We
consider the Zakharov-Shabat spectral problem \cite{zs} on the semi-line
\begin{equation}\label{zs-spectral}
x>0\quad:\quad\phi_x = ik[\phi,\sigma_3]+  
Q(x)\phi\ ,\quad Q=\left(\begin{array}{cc}0 & q \\ r & 0 \end{array}\right)
\end{equation}
with given boundary values
\begin{equation}\left.q(x)\right|_{x=0^+}=q_0\ ,\quad
\left.r(x)\right|_{x=0^+}=r_0\ .\end{equation}
Jost solutions \cite{levitan} can be defined as the solution of the integral 
equations
\begin{align}
\left(\begin{array}{c}\psi_{11}^+(k,x) \\ \psi_{21}^+(k,x)
\end{array}\right)&=
\left(\begin{array}{c}1\\0\end{array}\right)+
\left(\begin{array}{c}\int_{0}^{x}dy\   q(y) \psi_{21}^+(k,y) \\
      \int_{0}^{x}dy\  r(y) \psi_{11}^+(k,y) e^{2ik(x-y )}
       \end{array}\right)\ ,\label{int-psi+}\\                          
\left(\begin{array}{c}
\phi_{12}^+(k,x) \\ \phi_{22}^+(k,x)\end{array}\right)&=
\left(\begin{array}{c}0\\ 1\end{array}\right)+
\left(\begin{array}{c}
-\int_{x}^{\infty}dy\   q(y) \phi_{22}^+(k,y)e^{-2ik(x-y )}\\
     \int_{0}^{x}dy\  r(y) \phi_{12}^+(k,y)\end{array}\right)
\label{int-phi+}\ ,
\end{align}
\begin{align}
\left(\begin{array}{c}
\phi_{11}^-(k,x) \\ \phi_{21}^-(k,x)\end{array}\right)&=
\left(\begin{array}{c}1\\ 0\end{array}\right)+
\left(\begin{array}{c}\int_{0}^{x}dy\   q(y) \phi_{21}^-(k,y) \\
    - \int_{x}^{\infty}dy\  r(y) \phi_{11}^-(k,y) e^{2ik(x-y)}
       \end{array}\right)\label{int-phi-}\\ 
\left(\begin{array}{c}
\psi_{12}^-(k,x) \\ \psi_{22}^-(k,x) \end{array}\right)&=
\left(\begin{array}{c}0\\ 1\end{array}\right)+
\left(\begin{array}{c}\int_{0}^{x}dy\ q(y) \psi_{22}^-(k,y)e^{-2ik(x-y )}\\
     \int_{0}^{x}dy\  r(y) \psi_{12}^-(k,y) \end{array}\right).
\label{int-psi-}
\end{align}

\subsection{Boundary behaviors}

The so-called {\em physical solution} 
\begin{equation}\label{phys-sol}
\Phi(k,x)=(\phi_1^-,\phi_2^+)\ ,\end{equation}
 obeys the bounds
\begin{align}
&\left.\Phi(k,x)\right|_{x=0^+} =
\left(\begin{array}{cc}1 & \rho^+ \\ \rho^- & 1\end{array}\right)\ ,
\label{bound-phys-0}\\
&\left.\Phi(k,x)\right|_{x\to\infty}=
\left(\begin{array}{cc}\tau^- & 0\\ 0 &\tau^+\end{array}\right)\ ,
\label{bound-phys-inf}\end{align}
where the {\em reflection coefficients} $\rho^\pm(k)$ are defined by reading
the above integral equations at $x=0^+$, namely
\begin{align}\label{reflec}
&\rho^+(k)=-\int_{0}^{\infty}dy\   q(y) \phi_{22}^+(k,y)e^{2iky}
\notag\\
&\rho^-(k)=- \int_{0}^{\infty}dy\  r(y) \phi_{11}^-(k,y) e^{-2iky}\ ,
\end{align}
and the  {\em transmission coefficients} $\tau^\pm(k)$ by making the limit
as $x\to\infty$ i.e.
\begin{align}\label{transm}
&\tau^+(k)=1+\int_{0}^{\infty}dy\ r(y)\phi_{12}^+(k,y)\ ,
\notag\\
&\tau^-(k)=1+\int_{0}^{\infty}dy\ q(y)\phi_{21}^-(k,y)\ .
\end{align}
By direct computation with \eqref{zs-spectral} we demonstrate that the 
determinant of any solution is independent on $x$ and hence can be calculated
at $x=0$ or at $x\to\infty$. Applied to $\Phi$ it gives
\begin{equation}\label{unitarity}
1-\rho^+\rho^-=\tau^+\tau^-\ ,
\end{equation}
which is called the {\em unitarity relation}.

The rest of the solution, namely 
\begin{equation}\label{int-sol}
\Psi(k,x)=(\psi_1^+,\psi_2^-)\ ,\end{equation}
is  used to solve the inverse problem and we call it the {\em intermediate 
solution}. It obeys the following bounds for $k\in{\mathbb R}$
\begin{align}
&\left.\Psi(k,x)\right|_{x=0^+} =
\left(\begin{array}{cc}1 & 0 \\ 0 & 1\end{array}\right)\ ,
\label{bound-int-0}\\
&\left.\Psi(k,x)\right|_{x\to\infty}\sim
\left(\begin{array}{cc}1/\tau^+ & -e^{-2ikx}\rho^+/\tau^+\\
-e^{2ikx}\rho^-/\tau^-  & 1/\tau^-\end{array}\right)\ ,
\label{bound-int-inf}\end{align}
(the boundary behavior as $x\to\infty$ is actually obtained by using the
relations that follow).
The intermediate solution is related to the physical solution by
the following relations for $k\in{\mathbb R}$
\begin{align}\label{Hilbert}
&\psi_1^+ -\phi_1^-= -e^{2ikx}\rho^-\psi_2^-\ ,\notag\\
&\phi_2^+ - \psi_2^- =e^{-2ikx}\rho^+\psi_1^+\ ,
\end{align}
which can be proved in different ways, for instance by the
computation of the differences out of \eqref{int-phi+}\eqref{int-psi-} and the
comparison the resulting integral equation with the one for the r.h.s. of
\eqref{Hilbert}. A simpler but equivalent method is to express the vector
solution $\phi_1^-$ on the basis $\psi_1^+$ and $e^{2ikx}\psi_2^-$.
The coefficients of this expansion are then calculated by comparing the
boundary values in $x=0^+$.  Another proof goes with expressing the solution
$\Psi$ in terms of $\Phi$ as 
\begin{equation}\label{H-diff}
\Phi(k,x)=\Psi(k,x)\ e^{-ik\sigma_3x}
\left(\begin{array}{cc}1 & \rho^+(k) \\ \rho^-(k) & 1\end{array}\right)
e^{ik\sigma_3x}\ ,
\end{equation}
(computed by matching the values in  $x=0^+$) which in particular readily 
provides the behavior \eqref{bound-int-inf} from  \eqref{bound-phys-inf}.

\subsection{Analytical properties}

Using standard method \cite{levitan} one obtains that, for potentials vanishing
fast enough as $x\to\infty$, the vector $\psi_1^+$ is analytic in the upper
half $k$-plane and $\psi_2^-$ in the lower half. Next, the vector $\phi_2^+$ is
meromorphic function of $k$ in the upper half-plane with $N^+$ simple poles
$k_n^+$ which are the zeroes of the analytic function $1/\tau^+(k)$.  Indeed
from \eqref{bound-int-inf} we have 
\begin{equation}\label{un-sur-tau}
\frac1{\tau^+}=1+\int_0^\infty q\psi_{21}^+\ ,\end{equation}  
with $\psi_{21}^+$ analytic. Moreover the vector
\begin{equation}\tilde\phi_2^+=\frac1{\tau^+}\phi_2^+\end{equation} 
can be written from \eqref{int-phi+} as the solution of 
\begin{equation}
\left(\begin{array}{c} \tilde\phi_{12}^+(k,x) \\
\tilde\phi_{22}^+(k,x)\end{array}\right)= \left(\begin{array}{c}0\\
1\end{array}\right)+ \left(\begin{array}{c} \int_{0}^{x}dy\   q(y)
\tilde\phi_{22}^+(k,y)e^{-2ik(x-y )}\\ \int_{0}^{x}dy\  r(y)
\tilde\phi_{12}^+(k,y)\end{array}\right)\ , \end{equation} 
which shows that $\tilde\phi_2^+$ is analytic in $k$ and thus that
$\phi_2^+$ has poles where the analytic function $1/\tau^+(k)$ has zeroes.

Similarly, $\phi_1^-$ is meromorphic in the lower half-plane with $N^-$
poles $k_n^-$, zeroes of $1/\tau^-(k)$.  Consequently, from
their very definitions \eqref{reflec}\eqref{transm}, $\tau^\pm$ and $\rho^\pm$
have meromorphic extensions in their respective half-planes $\pm{\rm Im}(k)>0$
where they possess the $N^\pm$ simple poles $k_n^\pm$ (the bound states
locations).  In particular we have the relation, obtained simply by taking the
residues on \eqref{H-diff}
\begin{align}\label{residu+}
&\underset{k_n^+}{\rm Res}\{\phi_2^+\}=
\rho_n^+\psi_1^+(k_n^+)\exp[-2ik_n^+x]\ ,\quad
\rho_n^+ =\underset{k_n^+}{\rm Res}\{\rho^+\}\ ,\\
&\underset{k_n^-}{\rm Res}\{\phi_1^-\}=
\rho_n^-\psi_2^-(k_n^-)\exp[2ik_n^-x]\ ,\quad
\rho_n^-=\underset{k_n^-}{\rm Res}\{\rho^-\}\ .
\end{align}
The spectral data is then constituted by the set
\begin{equation}\label{S-data}
{\cal S}=\{ {\rm Im}(k)>0\ ,\ \rho^+(k)\ ;\ {\rm Im}(k)<0\ ,\ \rho^-(k)\}\ ,
\end{equation}
which is complete if we prove that it allows to reconstruct the potentials
$r(x)$ and $q(x)$ with $q(0^+)=q_0$ and $r(0^+)=r_0$.
 
\subsection{Large $k$ behaviors}

In order to solve the Hilbert problem \eqref{Hilbert} with the singularities
$k_n^\pm$ in the complex plane, one must calculate first the behaviour of the
solutions $\Phi(k,x)$ and $\Psi(k,x)$ as $|k|\to\infty$. Note that we follow
Mushkhelishvili \cite{mush} to refer to \eqref{Hilbert} as a Hilbert problem.

By using repeatidly the integration by part we obtain from the integral
equations \eqref{int-phi+}\eqref{int-psi-} the following regular asymptotic
behavior of the physical solution
\begin{align}\label{behav-phys}
&(\phi_1^-,\phi_2^+)={\bf 1}+\frac1{2ik}\left(\begin{array}{cc}
-\int_0^xrq & q \\ -r &\int_0^xrq\end{array}\right)\notag\\
&+\frac1{(2ik)^2}\left(\begin{array}{cc}
-\int_0^x[r_yq-rq\int_0^yrq] & -q_x+q\int_0^xrq \\ 
-r_x+r\int_0^xrq  & -\int_0^x[rq_y-rq\int_0^yrq]\end{array}\right)
+{\cal O}\left(\frac1{k^3}\right)\ ,
\end{align}
which has to be understood in ${\rm Im}(k)\le0$ for the first vector 
$\phi_1^-$ and in ${\rm Im}(k)\ge0$ for $\phi_2^+$.

The same procedure applied to the intermediate solution shows that it possess 
an essential singularity at $k=\infty$ on the real axis as indeed we get
\begin{align}\label{behav-inter}
&(\psi_1^+,\psi_2^-)={\bf 1}
+\frac1{2ik}\left(\begin{array}{cc}
-\int_0^xrq & q \\ -r &\int_0^xrq\end{array}\right)+\frac1{2ik}
\left(\begin{array}{cc}0 & -q_0e^{-2ikx} \\ r_0e^{2ikx} &0\end{array}\right)
\notag\\
&+\frac1{(2ik)^2}\left(\begin{array}{cc}
-\int_0^x[r_yq-rq\int_0^yrq]&
-q_x+q\int_0^xrq \\
-r_x+r\int_0^xrq
&-\int_0^x[rq_y-rq\int_0^yrq]\end{array}\right)\notag\\
&+\frac1{(2ik)^2}\left(\begin{array}{cc}
(qe^{2ikx}-q_0)r_0 &
(q_0 \int_0^xrq+q_0')e^{-2ikx}\\
(r_0 \int_0^xrq+r_0')e^{2ikx}
&(re^{-2ikx}-r_0)q_0\end{array}\right)+\cdots\ .
\end{align}
Consequently, when $q_0$ and $r_0$ do not vanish, the intermediate solution has
not the same behavior as the physical solution, in contrast with the infinite
line case. In particular the formulation of the solution of the Hilbert problem
will differ from the one of the infinite line.

Note for future use that from the definitions \eqref{reflec} of $\rho^\pm$ 
follow the large $k$ expansions
\begin{align}\label{exp-reflec}
\rho^+(k)=\frac1{2ik}\ q_0-\frac1{(2ik)^2}q_0'
+\frac1{(2ik)^3}(q_0''-q_0^2r_0)+{\cal O}(\frac1{k^4})\ ,\notag\\
\rho^-(k)=-\frac1{2ik}\ r_0-\frac1{(2ik)^2}r_0'
-\frac1{(2ik)^3}(r_0''-r_0^2q_0)+{\cal O}(\frac1{k^4})\ ,\end{align}
where we have defined
\begin{equation}q_0'=\left.\frac{\partial q(x)}{\partial x}\right|_{x=0^+}
\ ,\quad q_0''=\left.\frac{\partial^2q(x)}{\partial x^2}\right|_{x=0^+}\ ,
\end{equation}
so as for $r(x)$.

\section{Solution of the Hilbert problem \label{sec:hilb-pro}}

\subsection{Statement of the problem}

Let be given the set of {\em spectral data}
\begin{equation}
{\cal S}=\{ {\rm Im}(k)\ge 0\ ,\ \rho^+(k)\ ;\ {\rm Im}(k)\le 0\ ,
\ \rho^-(k)\}\ , \end{equation}
where $\rho^+(k)$ (resp. $\rho^-(k)$) is continuous and bounded on the real
axis and meromorphic in the upper (resp. lower) half plane with a finite number
of single poles $k_n^+$ (resp. $k_n^-$) and related residues $\rho_n^+$ (resp. 
$\rho_n^-$). Moreover they obey in their respective half-planes
\begin{equation}\label{bound-rho}
\rho^+(k)=\frac1{2ik}\ q_0+{\cal O}(\frac1{k^2})\ ,\quad
\rho^-(k)=-\frac1{2ik}\ r_0+{\cal O}(\frac1{k^2})
\end{equation}
where $q_0$ and $r_0$ are given numbers. The problem is to construct the two
matrices $\Phi(k,x)$ and $\Psi(k,x)$ solution of the 
Hilbert problem \eqref{Hilbert} obeying the following set of properties.

{\bf P.1 :}
Alternative expression of the Hilbert problem
$$\Phi=\Psi M\ ,\quad M=e^{-ik\sigma_3x}
\left(\begin{array}{cc} 1 & \rho^+\\ \rho^- & 1\end{array}\right)
e^{ik\sigma_3x} \ ,\quad x\ge0\ .$$

{\bf P.2 :}
The vector $\psi_1^+$ (resp. $\psi_2^-$) is analytic in ${\rm Im}(k)>0$
(resp. ${\rm Im}(k)<0$).
The vector $\phi_2^+$ (resp. $\phi_1^-$) is meromorphic in ${\rm Im}(k)>0$
(resp. ${\rm Im}(k)<0$) where it possess $N^+$ (resp. $N^-$) simple poles
$k_n^+$ (resp. $k_n^-$) with residues
$$\underset{k_n^+}{\rm Res}\{\phi_2^+\}=
\rho_n^+\psi_1^+(k_n^+)\exp[-2ik_n^+x]\ ,\quad
\underset{k_n^-}{\rm Res}\{\phi_1^-\}=
\rho_n^-\psi_2^-(k_n^-)\exp[2ik_n^-x]\ .$$

{\bf P.3 :}
The functions $\Psi(k,x)$ and $\Phi(k,x)$ obey the bounds
$$\Psi(k,0)=\left(\begin{array}{cc}1&0\\0&1\end{array}\right)\ ,\quad
\Phi(k,0)=\left(\begin{array}{cc}1&\rho^+(k)\\\rho^-(k)&1\end{array}\right)\ .
$$

{\bf P.4 :}
the function $\Psi(k,x)-{\bf 1}$ vanishes for $x<0$.

{\bf P.5 :}
the function $\Phi(k,x)$ has a Laurent expansion with a pole of zero order
at $k=\infty$:
$$\Phi(k,x)={\bf 1}+\frac1k\Phi^{(1)}(x)+{\cal O}(1/k^2)$$
(in the lower half-plane for the first vector, in the upper half-plane for the
second).

{\bf P.6 :}
the function $\Psi(k,x)$ has a Laurent expansion with an essential singularity
at $k=\infty$ on the real axis
$$\Psi={\bf 1}+\frac1k\Psi^{(1)}(x)+\frac1k\Psi_0^{(1)}e^{2ik\sigma_3x}
+\cdots$$

{\bf P.7 :}
There exists an off-diagonal matrix $Q(x)$ such that for $x>0$
$$\Psi_x=ik[\Psi,\sigma_3]+Q(x)\Psi\ ,$$
which is called the Zakharov-Shabat spectral equation.

{\bf P.8 :}
The function $Q$ obeys
$$\left.Q(x)\right|_{x=0^+}=
\left(\begin{array}{cc}0&q_0\\r_0&0\end{array}\right)\ ,\quad
\left.Q(x)\right|_{x=0^-}=0\ ,\quad 
Q(x)\underset{x\to\infty}{\longrightarrow} 0\ .$$

{\bf P.9 :}
The above defined function $Q(x)$ is also given by
$$Q(x)=i[\sigma_3,\Phi^{(1)}(x)]\ .$$

\subsection{The solution}

We prove now that the solution of the Hilbert problem \eqref{Hilbert} obeying
the properties P.1-P.9 above is given by the following equations. First
$\Phi(k,x)$ is given by the explicit expressions
\begin{align}
\phi_1^-(k,x)=&\left(\begin{array}{c}1\\ 0\end{array}\right)-\frac1{2i\pi}
\int_{-\infty+i0}^{+\infty+i0}\frac{d\lambda}{\lambda-k}\ 
\rho^-(\lambda)e^{2i\lambda x}\psi_2^-(\lambda,x)  \notag\\
&-\sum_{n=1}^{N^-}\frac{\rho_n^-}{k_n^--k}\psi_2^-(k_n^-,x)e^{2ik_n^-x}
\ ,\quad{\rm Im}(k)\le0
\label{cauchy-phi-1}\\
\phi_2^+(k,x)=&\left(\begin{array}{c}0\\ 1\end{array}\right)+\frac1{2i\pi}
\int_{-\infty-i0}^{+\infty-i0}\frac{d\lambda}{\lambda-k}\ 
\rho^+(\lambda)e^{-2i\lambda x}\psi_1^+(\lambda,x) \notag\\
&-\sum_{n=1}^{N^+}\frac{\rho_n^+}{k_n^+-k}\psi_1^+(k_n^+,x)e^{-2ik_n^+x}\
\ ,\quad{\rm Im}(k)\ge0 
\label{cauchy-phi-2}
\end{align}
where the function $\Psi(k,x)$ solves the following Cauchy-Green integral 
system of equations
\begin{align}
\psi_1^+(k,x)=&\left(\begin{array}{c}1\\ 0\end{array}\right)-\frac1{2i\pi}
\int_{-\infty-i0}^{+\infty-i0}\frac{d\lambda}{\lambda-k}\ 
\rho^-(\lambda)e^{2i\lambda x}\psi_2^-(\lambda,x)  \notag\\
&-\sum_{n=1}^{N^-}\frac{\rho_n^-}{k_n^--k}\psi_2^-(k_n^-,x)e^{2ik_n^-x}
\ ,\quad{\rm Im}(k)\ge 0
\label{cauchy-psi-1}\\
\psi_2^-(k,x)=&\left(\begin{array}{c}0\\ 1\end{array}\right)+\frac1{2i\pi}
\int_{-\infty+i0}^{+\infty+i0}\frac{d\lambda}{\lambda-k}\ 
\rho^+(\lambda)e^{-2i\lambda x}\psi_1^+(\lambda,x) \notag\\
&-\sum_{n=1}^{N^+}\frac{\rho_n^+}{k_n^+-k}\psi_1^+(k_n^+,x)e^{-2ik_n^+x}\
\ ,\quad{\rm Im}(k)\le 0 
\label{cauchy-psi-2}
\end{align}
In contrast with what occurs in the one dimensional case 
$x\in(-\infty,+\infty)$, here the integrals do not run on the real axis but
on contours along the real axis in the upper or lower half-plane
as indicated. We shall see that this apparently slight difference has a lot
of consequences and allows in particular to reconstruct the very solutions
$\Phi$ and $\Psi$ defined in the direct problem.

It is worth stressing here that we do not adress the problem of the {\em
complete characterization} of the spectral data $\rho^\pm(k)$, namely to find
the {\em necessary and sufficient} conditions which would ensure a unique
solution the the above Cauchy-Green intergral equations (such a problem has
been worked out in \cite{boutet}).  However we shall work with a set of
sufficient conditions on $\rho^\pm(k)$  ensuring that the reconstructed
potential does coincide with the one we started with. In particular for what
concerns solution of nonlinear evolutions, we will demonstrate that a set of
actual spectral data (constructed from a given potential at time $0$) evolve in
time such as to conserve this set of sufficient conditions.

We proceed now step by step to
demonstrate that the expressions \eqref{cauchy-phi-1}-\eqref{cauchy-psi-2} do
produce functions $\Phi$ and $\Psi$ that verify properties  P.1 to P.9.

\paragraph*{Property P.1 :}

For real values of $k$ the integral equations \eqref{cauchy-psi-1}
\eqref{cauchy-psi-2} can be written by virtue of Cauchy theorem
(by assumption $\rho^\pm(k)$ have no poles on the real axis)
\begin{align}
&\psi_1^+(k,x)=\left(\begin{array}{c}1\\ 0\end{array}\right)-\frac1{2i\pi}
\int_{-\infty+i0}^{+\infty+i0}\frac{d\lambda}{\lambda-k}\ 
\rho^-(\lambda)e^{2i\lambda x}\psi_2^-(\lambda,x)\notag\\
&-\rho^-(k)e^{2ikx}\psi_2^-(k,x) 
-\sum_{n=1}^{N^-}\frac{\rho_n^-}{k_n^--k}\psi_2^-(k_n^-,x)e^{2ik_n^-x}
\ ,\quad k\in{\mathbb R}\ ,
\label{cauchy-psi-1-reel}\end{align}
\begin{align}
&\psi_2^-(k,x)=\left(\begin{array}{c}0\\ 1\end{array}\right)+\frac1{2i\pi}
\int_{-\infty-i0}^{+\infty-i0}\frac{d\lambda}{\lambda-k}\ 
\rho^+(\lambda)e^{-2i\lambda x}\psi_1^+(\lambda,x)\notag\\
&-\rho^+(k)e^{-2ikx}\psi_1^+(k,x)
-\sum_{n=1}^{N^+}\frac{\rho_n^+}{k_n^+-k}\psi_1^+(k_n^+,x)e^{-2ik_n^+x}
\ ,\quad k\in{\mathbb R}\ .
\label{cauchy-psi-2-reel}\end{align}
Hence follow the Hilbert relations
\begin{equation}
\psi_1^+ -\phi_1^-= -e^{2ikx}\rho^-\psi_2^-\ ,\quad
\phi_2^+ - \psi_2^- =e^{-2ikx}\rho^+\psi_1^+\ ,
\end{equation}
and by rearrangement of the vectors into matrices, P.1 is proved.

\paragraph*{Property P.2 :} 
The analytical properties of the various vectors follow from their very
expressions. Indeed, as the set of single poles of $\Phi$ is explicitely
displayed, one has only to prove that the Cauchy-like integrals produce
holomorphic functions \cite{mush}. Such is the case if the integrant is a
continuous bounded function of $\lambda\in{\mathbb R}$ vanishing as
$\lambda\to\infty$. This actualy results from the assumed spectral data
$\rho^\pm(\lambda)$ and from the large $k\in{\mathbb R}$ properties of
$\Psi(k,x)$ deduced from the integral equations \eqref{cauchy-psi-1} and
\eqref{cauchy-psi-2} as the property P.6 that follows.

\paragraph*{Property P.3 :} 
The equations \eqref{cauchy-psi-1} \eqref{cauchy-psi-2} evaluated at $x=0$
($x$ is a parameter) lead to the system
\begin{align}
\psi_1^+(k,0)=&\left(\begin{array}{c}1\\ 0\end{array}\right)-\frac1{2i\pi}
\int_{-\infty-i0}^{+\infty-i0}\frac{d\lambda}{\lambda-k}\ 
\rho^-(\lambda)\psi_2^-(\lambda,0)\notag\\
&-\sum_{n=1}^{N^-}\frac{\rho_n^-}{k_n^--k}\psi_2^-(k_n^-,0)\ ,\\
\psi_2^-(k,0)=&\left(\begin{array}{c}0\\ 1\end{array}\right)+\frac1{2i\pi}
\int_{-\infty+i0}^{+\infty+i0}\frac{d\lambda}{\lambda-k}\ 
\rho^+(\lambda)\psi_1^+(\lambda,0)\notag\\
&-\sum_{n=1}^{N^+}\frac{\rho_n^+}{k_n^+-k}\psi_1^+(k_n^+,0)\ .
\end{align}
Now, as we assume a {\em unique solution} to the integral equations, it is 
sufficient to verify that $\Psi(k,0)={\bf 1}$ is indeed solution of the
above system. For $\psi_1^+(k,0)$ we have then the equation
$$\psi_1^+(k,0)=\left(\begin{array}{c}1\\ 0\end{array}\right)-\frac1{2i\pi}
\int_{-\infty-i0}^{+\infty-i0}\frac{d\lambda}{\lambda-k}\ 
\rho^-(\lambda)\left(\begin{array}{c}0\\1\end{array}\right)
-\sum_{n=1}^{N^-}\frac{\rho_n^-}{k_n^--k}
\left(\begin{array}{c}0\\1\end{array}\right)\ ,$$
which by contour integration in the lower half plane, and thanks to the 
properties of $\rho^-$, easily gives $\psi_1^+(k,0)=(1,0)^T$
(remember that $k$ lives in the upper half-plane). The same procedure
applies to the second vector and gives $\psi_2^-(k,0)=(0,1)^T$.

Next, by using exactly the same method with $\Phi(k,0)$, and taking care of the
pole $\lambda=k$, we do obtain the property P.3 (note that one could also use
P.1 to derive the boundary value of one of the functions from the other one).  

\paragraph*{Property P.4 :}

For $x<0$, thanks to the factor  $e^{2i\lambda x}$ appearing in the expression
\eqref{cauchy-psi-1}, we can perform the integration in the lower half-plane
for $\psi_1^+$. The contribution of the poles exactly cancels out and
$\psi_1^+(k,x)=(1,0)^T$ for $x<0$. The same procedure holds for $\psi_2^-$
which proves P.4. 

\paragraph*{Property P.5 :}

The expansion of $\Phi(k,x)$ at large $k$ readily results from the epressions
\eqref{cauchy-phi-1} and \eqref{cauchy-phi-2} where the Cauchy integral
behave well. Indeed, taking e.g. $\phi_1^-$, we integrate above the real axis
where, for $x>0$, the exponential is bounded.

We obtain in particular  for the matrix 
$\Phi^{(1)}(x)=(\phi_1^{(1)},\phi_2^{(1)})$
\begin{align}
&\phi_1^{(1)}=\frac1{2i\pi}\int_{-\infty+i0}^{+\infty+i0}
\rho^-(\lambda)e^{2i\lambda x}\psi_2^-(\lambda,x)
+\sum_{n=1}^{N^-}\rho_n^-\psi_2^-(k_n^-,x)e^{2ik_n^-x}\ ,\\
&\phi_2^{(1)}=-\frac1{2i\pi}
\int_{-\infty-i0}^{+\infty-i0}
\rho^+(\lambda)e^{-2i\lambda x}\psi_1^+(\lambda,x)
+\sum_{n=1}^{N^+}\rho_n^+\psi_1^+(k_n^+,x)e^{-2ik_n^+x}\ .
\end{align}

\paragraph*{Property P.6 :}

The expansion of $\Psi(k,x)$ at large $k\in{\mathbb R}$ cannot be obtained
directly from the integral equations \eqref{cauchy-psi-1} and
\eqref{cauchy-psi-2} (one sees on \eqref{cauchy-psi-1} that $\exp[2i\lambda x]$
is integrated for $x>0$ on a contour lying in the lower half $\lambda$-plane),
but rather form the versions \eqref{cauchy-psi-1-reel} and
\eqref{cauchy-psi-2-reel} where the Cauchy integrals behave well. Then the same
procedure as for P.5 applies to get the expansion given in P.6 (it is actually
easier to use P.5 and P.1 together with the expansion of the spectral data
$\rho^\pm(k)$ to demonstrate P.6). We obtain in particular, according to the
notations in P.6,
\begin{equation}
\Psi^{(1)}(x)=\Phi^{(1)}(x)\ ,\quad 
\Psi^{(1)}_0=\left(\begin{array}{cc} 0&-\frac{q_0}{2i}\\
\frac{r_0}{2i}&0\end{array}\right)\ ,
\end{equation}
which matches the expressions \eqref{behav-inter} (and can be pursued to next
orders).

\paragraph*{Property P.7 :}

This is a fundamental property in the process of solving the inverse problem as
indeed it gives the expression of the potential from the solution of the
Cauchy-Green integral equations \eqref{cauchy-psi-1}\eqref{cauchy-psi-2}, in
other words in terms of the spectral data $\rho^\pm(k)$.

To perform the proof of P.7 it is first more convenient to rewrite the
integral equations as follows
\begin{align}
&\psi_1^+=\left(\begin{array}{c}1\\ 0\end{array}\right)
-\frac1{2i\pi}\int_{C_-}\frac{d\lambda}{\lambda-k}
\rho^-e^{2i\lambda x}\psi_2^-  \ ,
\label{cauchy-1}\\
&\psi_2^-=\left(\begin{array}{c}0\\ 1\end{array}\right)
+\frac1{2i\pi}\int_{C_+}\frac{d\lambda}{\lambda-k}
\rho^+e^{-2i\lambda x}\psi_1^+\ ,
\label{cauchy-2}
\end{align}
where $C_+$ (resp. $C_-$) is a smooth contour in the upper (resp. lower)
half-plane extending from $-\infty$ to $\infty$ and passing {\em over} all
poles $k_n^+$ (resp. {\em under} all poles $k_n^-$). Note that $\psi_1^+$ is
written for ${\rm Im}(k)>0$ and hence there is no contribution of the pole
$\lambda=k$ in the lower half-plane. The same holds reversely for $\psi_2^-$. 
In short these integral equations read
\begin{equation}\label{cauchy-psi}
\Psi(k,x)={\bf 1}+\frac1{2i\pi}\int_{\cal C}\frac{d\lambda}{\lambda-k}
\Psi(\lambda,x)R(\lambda,x)\ ,\end{equation}
where
\begin{equation}\label{R}
R(k,x)=e^{-ik\sigma_3x}
\left(\begin{array}{cc}0&\rho^+(k)\\-\rho^-(k)&0\end{array}\right)
e^{ik\sigma_3x}\ ,\end{equation}
and where the contour of integration has to be understood as being $C_-$ for 
the first vector of the product $\Psi R$, and $C_+$ for the second.

Thanks to the particular dependence in $x$ of $R(k,x)$ it is easy to compute
$\Psi_x+ik[\sigma_3,\Psi]$ and by algebraic manipulations to get
\begin{align}\label{inteq-zs}
\Psi_x+ik[\sigma_3,\Psi]=&-\frac1{2\pi}\int_{\cal C}d\lambda[\sigma_3,\Psi R]
\notag\\
&+\frac1{2i\pi}\int_{\cal C}\frac{d\lambda}{\lambda-k}
\left(\Psi_x+i\lambda[\sigma_3,\Psi]\right)R\ .\end{align}
Then by defining the off-diagonal matrix
\begin{equation}\label{Q}
Q(x)=-\frac1{2\pi}\int_{\cal C}d\lambda[\sigma_3,\Psi(\lambda,x) R(\lambda,x)]
\ ,\end{equation}
we can compare the integral equation \eqref{inteq-zs} to the integral equation
for the matrix $Q(x)\Psi(\lambda,x)$. As the solution is assumed to be unique
we arrive at the desired result
\begin{equation}\Psi_x+ik[\sigma_3,\Psi]=Q\Psi\ .\end{equation}

\paragraph*{Property P.8 :}

We derive now from the above definition of $Q(x)$ the properties P.8 and to 
that end rewrite \eqref{Q} as
\begin{align}\label{qr}
&q(x)=-\frac1\pi\int_{C_+}\rho^+(\lambda)\psi_{11}^+(\lambda,x)e^{-2i\lambda x}
\ ,\notag\\
&r(x)=-\frac1\pi\int_{C_-}\rho^-(\lambda)\psi_{22}^-(\lambda,x)e^{2i\lambda x}
\ .\end{align}
First, for $x<0$, we can close the contours with the circle of infinite radius
in the related half-planes and readily obtain  that $q$ and $r$ vanish
(remember that the contours pass beyond the poles of $\rho$).

To evaluate the boundary values in $x=0$ we use $\psi_{11}^+(k,0)=
\psi_{22}^-(k,0)=1$ (the function $\Psi$ is continuous in $x=0$), but keep the 
exponentials as they produce discontinuous functions in $x=0$. Then with help
of  the large $k$ behaviors \eqref{bound-rho} and the formulae
\begin{equation}\label{distrib}
x>0\ \Rightarrow\ -\frac1{2i\pi}\int_{C_+}\frac{e^{-2i\lambda x}}{\lambda}
=\frac1{2i\pi}\int_{C_-}\frac{e^{2i\lambda x}}{\lambda}=1\ ,
\end{equation}
we do obtain $q(0^+)=q_0$ and $r(0^+)=r_0$, where $q_0$ and $r_0$ are given
from the spectral data $\rho^\pm$ in \eqref{bound-rho}. 

Note that the procedure can be applied to compute the $x$-derivative of the
potential by first regularizing it in $x=0$ and by using the next order
expansion of $\rho(k)$ given in \eqref{exp-reflec}. Indeed we have
\begin{equation}
\frac{\partial}{\partial x}\left\{q(x)-q_0\theta(x)\right\}=-\frac1\pi
\frac{\partial}{\partial x}
\int_{C_+}[\rho^+\psi_{11}^+ -\frac{q_0}{2i\lambda}]
e^{-2i\lambda x}\ ,\end{equation}
and applying the same procedure we obtain eventually
\begin{equation}
\lim_{x\to0^+}\left\{ \frac{\partial}{\partial x}q(x)\right\}=
\left.\frac{\partial}{\partial x}\left\{q(x)-q_0\theta(x)\right\}\right|_{x=0}
=q_0'\ .\end{equation}

The large $x$ vanishing behaviors of $q$ and $r$ follow from 
their expressions \eqref{qr} rewritten as (we consider only $q(x)$, the same
computation works as well for $r(x)$)
\begin{align}\label{q(x)}
q(x)=&-\frac1\pi\int_{-\infty+i0}^{+\infty+i0}
\rho^+(\lambda)\psi_{11}^+(\lambda,x)e^{-2i\lambda x}\notag\\
&+2i\sum_{n=1}^{N^+}\rho_n^+\psi_{11}^+(k_n^+,x)e^{-2ik_n^+x}
\ .\end{align}
The above expression vanishes at large $x$ because the path integral
vanishes and because 
\begin{equation}
\psi_{11}^+(k,x)\underset{x\to\infty}{\longrightarrow}\frac1{\tau^+(k)}\ ,
\quad \left.\frac1{\tau^+(k)}\right|_{k=k_n^+}=0\ ,\end{equation}
which results from \eqref{bound-int-inf} and from the very definition of the
poles $k_n^+$.

\paragraph*{Property P.9 }
results from the fact that, due to P.1 and P.7, the function $\Phi$ also
solves Zakharov-Shabat, i.e.
\begin{equation}\label{lax-x}
\Phi_x=ik[\Phi,\sigma_3]+Q(x)\Phi\ .\end{equation}
Then by inserting in it the large $k$ behavior given by P.5 we must arrive at
p.9. To check consistency we have then to demonstrate that the following two
expressions of $q(x)$
\begin{align}
&-\frac1\pi\int_{-\infty+i0}^{+\infty+i0}
\rho^+(\lambda)\psi_{11}^+(\lambda,x)e^{-2i\lambda x}
+2i\sum_{n=1}^{N^+}\rho_n^+\psi_{11}^+(k_n^+,x)e^{-2ik_n^+x}\ ,\\
&-\frac1\pi\int_{-\infty-i0}^{+\infty-i0}
\rho^+(\lambda)\psi_{11}^+(\lambda,x)e^{-2i\lambda x}
+2i\sum_{n=1}^{N^+}\rho_n^+\psi_{11}^+(k_n^+,x)e^{-2ik_n^+x}\ ,
\end{align}
do coincide. This simply results from the starting hypothesis of a spectral
transform $\rho(k)$ which is continuous and bounded on the real axis (no
poles).  Then both path integral hereabove are equal. 

Note that the relation P.9 giving $Q(x)$ can be evaluated in $x=0$, which does
lead to the expected result thanks to the boundary behavior
\eqref{bound-phys-0} of $\Phi$ in $x=0$ and the large $k$ behaviors of $\rho^+$
and $\rho^-$. Note also that, in order to compute $q(0^+)$ one should first
deform the contour to pass {\em above} all poles $k_n^+$ and then make use of
\eqref{distrib}. 

\subsection{Reductions}

The 3 particular evolutions that we shall study then result from a 
{\em reduction} from the two potentials $r$ and $q$ to only one. 
The first reduction concerns SRS and reads
\begin{equation}\label{red-q}
\overline Q=\sigma_2\ Q\ \sigma_2\ \Leftrightarrow r=-\bar q\ .
\end{equation}
Then by direct computation one easily proves that
$\sigma_2\overline{\Psi(\overline k,x)}\sigma_2$ solves the same
spectral equation  \eqref{zs-spectral} as $\Psi(k,x)$ and possess the same
boundary behavior \eqref{bound-int-0} in $x=0$. 
Those two functions then coincide,
\begin{equation}
\Psi(k,x)=\sigma_2\overline{\Psi(\overline k,x)}\sigma_2\ , \end{equation}
and examination of their boundary behavior as $x\to\infty$ gives
the reduction constraint on the spectral transform
\begin{equation}\label{red-spectr-srs}
\overline{\tau^-(\overline k)}=\tau^+(k)\ ,\quad
\overline{\rho^-(\overline k)}=-\rho^+(k)\ .\end{equation}
Considering the poles and residues, the above expressions imply
\begin{equation}\label{reduc-poles}
N^+=N^-\ ,\quad \overline{\rho_n^-}=-\rho_n^+\ ,\quad 
\overline{k_n^-}=k_n^+\ .\end{equation}

The second case is related to SG in thge light cone, it reads
\begin{equation}\label{red-q-sg}
Q=\sigma_2Q\sigma_2\ \Leftrightarrow r=- q\ .
\end{equation}
In that case it is easy to prove that
\begin{equation}\label{red-jost-sg}
\Psi(k,x)=\sigma_2\Psi(-k,x)\sigma_2\ , \end{equation}
and examination of the boundary behavior as $x\to\infty$ and $x\to 0$ gives
the reduction constraint on the spectral transform
\begin{align}\label{red-spectr-sg}
&\rho^+(k,t)=-\rho^-(-k,t)\ ,\quad\tau^-(-k)=\tau^+(k)\ ,\notag\\
&N^+=N^-\ ,\quad \rho_n^-=-\rho_n^+\ .\end{align}

In the case of sine-Gordon, we also need to implement the reality constraint
$q\in{\mathbb R}$ for which both relations \eqref{red-q} and \eqref{red-q-sg}
have to hold together. As a consequence, by direct manipulations of relations
\eqref{red-spectr-srs} and \eqref{red-spectr-sg}, the resulting constraint
on the spectral transform $\rho=\rho^+$ which comes in addition to 
\eqref{red-spectr-sg} reads
\begin{equation}\label{real-red-sg}
\overline\rho(-\overline k)=\rho(k)\ .
\end{equation}
As a consequence the poles  in the complex plane come by pair, namely if
$k_n$ is a pole of $\rho(k)$ with residue $\rho_n$, then $-\overline k_n$
is also a pole with residue $-\overline \rho_n$. In short
\begin{equation}\label{real-red-sg-poles}
\underset{k_n}{\rm Res}\{\rho(k)\}=\rho_n\ \Rightarrow\ 
\underset{-\overline k_n}{\rm Res}\{\rho(k)\}=-\overline\rho_n\ .
\end{equation}

\section{The Dirichlet problem for SRS\label{sec:srs}}
\subsection{Boundary values and Lax pair}

The formalism of the direct and inverse problem for the Zakharov-Shabat system
on the semi-line $x>0$ is now applied to the {\em Stimulated Raman Scattering}
(SRS) equations for the pulse envelopes of the pump $a(k,x,t)$, the Stokes
$b(k,x,t)$ and the medium excitation $q(x,t)$
\begin{align}\label{srs}
&\partial_xa=qbe^{2ikx}\ ,\quad \partial_xb=-\overline q ae^{-2ikx}\notag\\
&\partial_tq =-\int g(k)dk\ a\overline{b}e^{-2ikx}\ ,\end{align}
where $g(k)$ measures the coupling (related to Raman gain) and the parameter
$k$ is a missmatch wave number resulting from group velocity dispersion
\cite{js}. The Lax pair for SRS has been written in the case
$g(k)=\delta(k-k_0)$ in \cite{chusco} and regularized by means of the parameter
$k$ in \cite{jl} which found its physical meaning in \cite{js}.

Dirichlet conditions are prescribed on the line $x=0\ ,\ t>0$, 
\begin{equation}\label{cond-x=0}
\left.a(k,x,t)\right|_{x=0}=A(t)\ ,\quad 
\left.b(k,x,t)\right|_{x=0}=B(t)\ ,\end{equation}
and on the line $x>0\ ,\ t=0$,
\begin{equation}\label{cond-t=0}
\left.q(x,t)\right|_{t=0}=c(x)\ .\end{equation}
Note that we could assume boundary values $A(k,t)$ and $B(k,t)$ instead 
of $A(t)$ and $B(t)$, but here we have chosen to leave the $k$-dependence in
the arbitrary distribution $g(k)$. This is equivalent to assuming separable
dependence of $A$ an $B$ in $k$ and $t$.
The given boundary value $c(x)$  is assumed to be of Schwartz 
type (it decreases at large $x$ with all its derivatives faster that any
polynomial) but the physically relevant case is a medium initially at rest,
namely $c(x)=0$. Note that the compatibility of this set of Dirichlet
boundary data is infered from the system \eqref{srs} which gives
\begin{equation}
\left.\partial_xa(k,x,0)\right|_{x=0}=c(0)B(0)\ ,\quad
\left.\partial_xb(k,x,0)\right|_{x=0}=-\overline c(0)A(0)\ .\end{equation}

The above Dirichlet problem has been solved in \cite{js} and, for sake of
completeness, we give hereafter the main lines of the derivation of the
solution.  The vector $(a,be^{2ikx})^T$ is expanded on the  basis $\psi_1^+$
and $\phi_2^+e^{2ikx}$, the coefficients being obtained by examination of the
values in $x=0$. The result reads
\begin{equation}
\left(\begin{array}{c}a\\ be^{2ikx}\end{array}\right)=
(A-\rho B)\psi_1^+ + B\phi_2^+\ .
\end{equation}
As a consequence we have
\begin{equation}
\left(\begin{array}{c}a\\ b\end{array}\right)
\underset{x\to\infty}{\longrightarrow}
\left(\begin{array}{c}(A-\rho B)/\tau\\ 
\overline\rho{}(A-\rho B)/\overline\tau+ B\tau\end{array}\right)\ .
\end{equation}

The time dependence is plug into the spectral data by requiring that the 
solution  $\mu^+=(\psi_1^+,\phi_2^+)$ of the Zakharov-Shabat equation 
be also solution for $k\in{\mathbb R}$ of
\begin{align}\label{lax-t}
&\mu^+_t=V^+\mu^+ +\mu^+ e^{-ik\sigma_3x}\Omega^+ e^{ik\sigma_3x}\ ,\\
&V^+=\frac{1}{4i}\int\frac{g(\lambda)d\lambda}{\lambda-(k+i0)}
\left(\begin{array}{cc}|a|^2-|b|^2 & 2a\overline be^{-2i\lambda x}\\
2\overline abe^{2i\lambda x}   &|b|^2-|a|^2\end{array}\right)\ .
\notag\end{align}
Note that, as the matrix $V(k,x,t)$ is discontinuous when $k$ crosses the
real axis, we have to chose a given half-plane.
A similar equation could be written for $\mu^-=(\phi_1^-,\psi_2^-)$ with $V^-$
and $\Omega^-$ but, due to the reduction symmetry, it is redundant.  The
compatibility between the two operators \eqref{lax-x}\eqref{lax-t} with the
constraint $\Omega_x=0$, results in the equation \eqref{srs} within the
reduction \eqref{red-q}.  

\subsection{Evolution of the spectral transform}

The free entry $\Omega^+=\Omega^+(k,t)$ is the {\em dispersion relation} and,
according to the general theory \cite{jl}, it is stricly dependent of the
choice of the solution.  It is called dispersion relation because, in the
infinite line case, it coincides with the dispersion law of the linearized
evolution. The differential equation \eqref{lax-t} is evaluated at $x=0$ and
$x\to\infty$. This provides 8 equations for the 4 components of $\Omega^+(k,t)$
and the 4 evolutions equations of $\rho(k,t)$, $\tau(k,t)$ and their
conjugates (these evolutions are compatible which actually results from the
reduction). We obtain 
\begin{equation}
\Omega^+(k,t)=-\frac{i}4{\cal I}(k)
\left[2\rho(k,t)\overline AB\sigma_3+\left(
\begin{array}{cc}|A|^2-|B|^2&0\\2\overline AB &-|A|^2+|B|^2\end{array}\right)
\right]\ ,\end{equation}
(note that the dispersion relation depends on the spectral transform 
itself), and the Riccati evolution
\begin{equation}\label{evol-rho-srs}
\rho_t=\frac{i}2{\cal I}
\left[\overline AB\ \rho^2-(|A|^2-|B|^2)\ \rho-A\overline B\right]\ ,
\end{equation}
where we have defined
\begin{equation}{\cal I}(k)=\int\frac{g(\lambda)d\lambda}{\lambda-k}
\ ,\quad {\rm Im}(k)>0\ .
\end{equation}

The main fact is that the above evolution conserves the Laurent expansion
of $\rho(k,t)$. Indeed, seeking a solution of \eqref{evol-rho-srs} as the
power series
\begin{equation}\label{laurent}
\rho(k,t)=\sum_{j=1}^p \rho_j(t)k^j+\sum_{n=0}^\infty
\frac1{k^n}\rho^{(n)}(t)\ ,\end{equation}
allows to prove by induction that $\rho_j(t)=0$ for $j\geq 2$. For $j=1$ we 
have the following evolution equation
\begin{equation}
\frac{\rho_{1t}}{\rho_1^2}=\frac i2A\overline B\int d\lambda\ g(\lambda)
=\alpha(t)\ ,
\end{equation}
which is solved to get
\begin{equation}
\rho_1(t)=\frac{\rho_1(0)}{1-\rho_1(0)\int_0^t\alpha(t)}\ .
\end{equation}
Then if $\rho_1$ vanishes at time $t=0$, it vanishes for all $t$.
Such  is also the case for $\rho^{(0)}$ whose
evolution (with  $\rho_1(t)=0$) simply reads
\begin{equation}
\rho^{(0)}_{t}=0\ .
\end{equation}
Consequently, as $\rho_1$ and $\rho^{(0)}$ vanish at initial time (because
$\rho(k,0)$ is the spectral transform of the Dirichlet datum $q(x,0)$),  they
vanish for all time and the solution of the Riccati equation 
(\ref{evol-rho-srs}) possess the right behavior for large $k$, compatible with
(\ref{exp-reflec}).

The coefficient of 
$k^{-1}$ gives rise to
\begin{equation}\label{diff-rho1}
\partial_t\rho^{(1)}(t)=\frac i2A\overline B\int d\lambda\ g(\lambda)\ .
\end{equation}
The evolution \eqref{srs} written in $x=0$ gives
\begin{equation}
\partial_tq_0(t)=-A\overline B\int d\lambda\ g(\lambda)\ ,
\end{equation}
and moreover we have, by construction,
\begin{equation}\rho^{(1)}(0)=\frac{q_0(0)}{2i}\ .\end{equation}
Consequently the unique solution to \eqref{diff-rho1} is
\begin{equation}\label{check-k}
\rho^{(1)}(t)=\frac{q_0(t)}{2i}\end{equation}
as required by \eqref{exp-reflec}. 

The time evolution \eqref{evol-rho-srs} of the spectral transform has 
coefficients depending only on the input boundary values $A(t)$ and $B(t)$
and contains all the relevant information on the nonlinear scattering of the
three waves $q$, $a$ and $b$. In some cases this Riccati equation is
explicitely solvable, for instance when $A(t)$ and $B(t)$ are proportionnal
(which is physically meaningful) as illustrated in \cite{js}. It can then
be checked on the explicit expression of $\rho(k,t)$ of \cite{js} that it
does obey the good large $k$ asymptotics. The fact of being able to provide
an explicit expression of the spectral transform in a class of  physically
interesting cases makes the very interest of the present approach of
boundary value problems.

\subsection{Why it works}

The Dirichlet open-end problem is well posed if the solution $q(x,t)$
constructed hereabove is proved to obey for all $t$ the boundary value
\begin{equation}
\left.q(x,t)\right|_{x=0^+}=q_0(t)\ ,\end{equation}
and to belong to the class of functions such that
\begin{equation}
\lim_{x\to\infty}q(x,t)=0\ .\end{equation}
This proof goes along the following steps.

\paragraph{Step 1:} the solution $\rho(k,t)$ of the Riccati equation
\eqref{evol-rho-srs} possess the large $k$ expansion 
\begin{equation}\rho(k,t)=\frac{q_0(t)}{2ik}+{\cal O}(1/k^2)\end{equation}
resulting from \eqref{check-k}. 
This result is central to the theory as we see here that the
time evolution of the spectral transform conserves the properties it has at
$t=0$, see \eqref{exp-reflec}. Note that the expansion can be pursued at any
order.

\paragraph{Step 2:} with such a behavior of $\rho(k,t)$ the property
 P.3 still holds, namely
\begin{equation}
\Psi(k,0,t)=\left(\begin{array}{cc}1&0\\0&1\end{array}\right)\ .\end{equation}
The proof of the above boundary value is obtained by following exactly the
method of section \ref{sec:hilb-pro} where $q_0$ has simply to be
understood as $q_0(t)$.

\paragraph{Step 3:} compute the limit of  $q(x,t)$ 
as $x\to0^+$ out of the expresion \eqref{q(x)} by following the
method of section \ref{sec:hilb-pro} used to prove property P.8.

\paragraph{Step 4:} compute the limit of  $q(x,t)$ 
as $x\to\infty$ out of the expresion \eqref{q(x)} again by following
the proof of property P.8. Note that the essential property to be used
is the fact that $\rho(k,t)$ is continuous and bounded for $k\in{\mathbb R}$
at any given $t$. Particular attention is required when, as $t$ evolves,
a pole of $\rho$ is {\em ``seen''} to cross the real axis. It has been 
demonstrated in \cite{bclp} that the process is continuous, or put in other
words, a resonance (pole in the lower half-plane) dissapears and a pole
appears while never belonging to the real axis. The same situation arises
for the Schr\"odinger eigenvalue problem \cite{hug2}.

\subsection{Summary of the method of solution of SRS}

The solution of SRS with  Dirichlet condition  is achieved along the
following steps:

1 - The spectral transform at time zero $\rho(k,0)$ is obtained form the
Dirichlet condition $\tilde q(x)$ by finding the vector solution 
$\phi_2(k,x,0)$ of the integral equations \eqref{int-phi+} in the upper half 
$k$-plane, namely
\begin{align}
&\phi_{12}^+(k,x,0) =
-\int_{x}^{\infty}dy\ \tilde q(y) \phi_{22}^+(k,y,0)e^{-2ik(x-y )}
\notag\\
&\phi_{22}^+(k,x,0)=1-\int_{0}^{x}dy\  \overline{\tilde q(y)} 
\phi_{12}^+(k,y,0)\ ,
\end{align}
and then $\rho(k,0)$ results from
\begin{equation}
\rho(k,0)=\left.\phi_{12}^+(k,x,0)\right|_{x=0^+}\ .\end{equation}

2 - $\rho(k,t)$ is calculated in the upper half $k$-plane from $\rho(k,0)$ by 
solving the Riccati time-evolution 
\begin{equation}\label{riccati-srs}
\rho_t=\frac{i}2{\cal I}
\left[\ \overline AB\ \rho^2-(|A|^2-|B|^2)\ \rho-A\overline B\ \right]\ ,
\end{equation}
with entries $A(t)$ and $B(t)$. The $N$ poles $k_n$ in the upper
half-plane and related residues $\rho_n$ of $\rho(k,t)$ must be identified for
each value of the parameter $t$.

3 - The integral system \eqref{cauchy-psi-1}\eqref{cauchy-psi-2}
which, within the reduction \eqref{red-spectr-srs}, simplifies to
\begin{align}
\psi_{11}^+(k,x,t)=&1+\frac1{2i\pi}
\int_{-\infty-i0}^{+\infty-i0}\frac{d\lambda}{\lambda-k}\ 
\overline{\rho(\overline\lambda,t)}e^{2i\lambda x}\psi_{12}^-(\lambda,x,t)
\notag\\
&+\sum_{n=1}^{N(t)}\frac{\overline{\rho_n(t)}}{\overline{k_n(t)}-k}
\psi_{12}^-(\overline{k_n(t)},x,t)e^{2i\overline{k_n(t)}x}\ ,
\end{align}
\begin{align}
\psi_{12}^-(k,x,t)=&\frac1{2i\pi}
\int_{-\infty+i0}^{+\infty+i0}\frac{d\lambda}{\lambda-k}\ 
\rho(\lambda,t)e^{-2i\lambda x}\psi_{11}^+(\lambda,x,t) \notag\\
&-\sum_{n=1}^{N(t)}\frac{\rho_n(t)}{k_n(t)-k}\psi_{11}^+(k_n,x,t)e^{-2ik_n(t)x}
\ ,\end{align}
is solved to get $\psi_{11}^+(k,x,t)$.  This system of equation holds for any
fixed value of the {\em ``parameters''} $x$ and $t$.

4 - Finally the solution $q(x,t)$ is obtained from expression \eqref{q(x)}, i.e.
\begin{align}
q(x,t)=&-\frac1\pi\int_{-\infty+i0}^{+\infty+i0}d\lambda\ 
\rho(\lambda,t)\psi_{11}^+(\lambda,x,t)e^{-2i\lambda x}\notag\\
&+2i\sum_{n=1}^{N(t)}\rho_n(t)\psi_{11}^+(k_n(t),x,t)e^{-2ik_n(t)x}\ .
\end{align}

\section{The Dirichlet problem for sine-Gordon\label{sec:sg}}
\subsection{Boundary values and Lax pair}

Another case where the method applies sucessfully is the
sine-Gordon equation in the light cone
\begin{equation}
\theta_{xt}+\sin\theta=0\ ,\end{equation}
which is solved here for Dirichlet conditions $\theta_0(t)$ and 
$\tilde\theta(x)$, namely
\begin{equation}
t\in[0,T]\ :\ \theta(0,t)=\theta_0(t)\ ,\quad 
x\in[0,\infty)\ :\ \theta(x,0)=\tilde\theta(x)\ ,\end{equation}
in the class of functions vanishing (modulo $2\pi$) with all derivatives as
$x\to\infty$ for every value of time $t$ (the value $T$ is arbitrary).

This equation is fundamental in many areas of physics and has been consequently
widely studied. The solution of the boundary value problem on the quadrant
also has been the subject of intense studies, see e.g.
\cite{bik-tar,beloko,fok-its,fokas,bea} and the references therein.

The Lax pair, found in \cite{lax-sg},
is given by the Zakharov-Shabat equation \label{zs} for a 
particular choice of potentials $r$ and $q$, namely 
\begin{equation}\label{lax-x-sg}
\Phi_x = ik[\Phi,\sigma_3]+  Q\Phi\ ,\quad 
Q= -\frac i2\sigma_2\theta_x \ ,
\end{equation}
together with the following evolution
\begin{equation}\label{lax-t-sg}
\Phi_t =-\frac i{4k}\left(\sigma_3\cos\theta+\sigma_1\sin\theta\right)\Phi
+\Phi e^{-ik\sigma_3x}\Omega e^{ik\sigma_3x}\ .
\end{equation}
We have chosen to write this Lax pair for the particular solution $\Phi$
defined in section \ref{sec:dir-pro}. Note that the above particular structure 
of the potential $Q$ obeys the reduction constraint \eqref{red-q-sg}. 

\subsection{Evolution of the spectral transform}

By using the behaviors \eqref{bound-phys-0} and \eqref{bound-phys-inf}
in \eqref{lax-t-sg} we calculate $\Omega$ and get the time
evolution of the spectral transform. Within the reduction
relations \eqref{red-spectr-sg} we eventually obtain
\begin{equation}\label{omega-sg}
\Omega=-\frac{i}{4k}\left(
\begin{array}{cc}\cos\theta_0-\rho(-k,t)\sin\theta_0&0\\
0&-\cos\theta_0+\rho(k,t)\sin\theta_0\end{array}\right)\ ,\end{equation}
\begin{equation}\label{evol-rho-sg}
\rho_t= \frac{i}{4k}\left[\rho^2\sin\theta_0-2\rho\cos\theta_0
-\sin\theta_0\right]\ .\end{equation}
Note that the reduction constraint $\overline\rho(-\overline k)=\rho(k)$
which ensures real value potential $\theta(x,t)$ is conserved by the above time
evolution (as soon as $\theta_0\in\mathbb R$ of course).

The evolution of $\rho(k,t)$ is of Riccati type and depends exclusively on the
boundary value $\theta_0(t)$, while the other Dirichlet condition
$\tilde\theta(x)$ fixes the initial value $\rho(k,0)$. Following exactly the
same method as for SRS we show that, as $\rho_1(0)$ and $\rho^{(0)}(0)$ vanish
(by construction), the large $k$ asymptotic behavior of the solution
$\rho(k,t)$ reads
\begin{equation}\label{check-k-sg} 
\rho(k,t)=\frac1k\rho^{(1)}+{\cal O}(1/k^2)\ ,\quad
\partial_t\rho^{(1)}=-\frac i4\sin\theta_0\ .\end{equation}
Thanks to the sine-Gordon equation evaluated in $x=0$ we can integrate the 
above time evolution and obtain
\begin{equation} \rho^{(1)}(t)=-\frac1{4i}\theta_x(0,t)=\frac{q_0(t)}{2i}
\ ,\end{equation}
as indeed, by construction of $\rho(k,0)$ from $\tilde\theta(x)$, we have
\begin{equation} \rho^{(1)}(0)=-\frac1{4i}\left.\tilde\theta_x\right|_{x=0}
=-\frac1{4i}\theta_x(0,0)\ .\end{equation}

The last point to consider is the apparent singularity of $\rho(k,t)$ when
$k\to0$ by real values. Actually for $k\to0$ the spectral transform possess
the following regular behavior 
\begin{equation}
\rho(k,t)=r_0(t)+kr_1(t)+{\cal O}(k^2)\ ,\end{equation}
that can be deduced from the very definition \eqref{reflec} of $\rho(k,t)$
(or equivalently by \eqref{rho-equiv} below), by expanding these expressions in
power of $k$. Inserting the above expansion in \eqref{evol-rho-sg} gives
the following unique solution vanishing for $\theta_0=0$,
\begin{equation}
\rho(0,t)=r_0(t)=\frac{\cos\theta_0-1}{\sin\theta_0}\ .
\end{equation}
The other terms $r_n(t)$ in the expansion are then recursively given in terms 
of $r_j$, ($j<n$), without constraint.

\subsection{Why it works}

The Dirichlet open-end problem is well posed if the solution $\theta(x,t)$
constructed hereabove is proved to obey for all $t$ the boundary value
\begin{equation}
\left.\theta(x,t)\right|_{x=0^+}=\theta_0(t)\ ,\end{equation}
and to belong to the class of functions such that
\begin{equation}
\lim_{x\to\infty}\theta_x(x,t)=0\ .\end{equation}
This proof goes along the following steps.

\paragraph{Step 1:} the solution $\rho(k,t)$ of the Riccati equation
\eqref{evol-rho-sg} possess the large $k$ expansion 
\begin{equation}\rho(k,t)=\frac{q_0(t)}{2ik}+{\cal O}(1/k^2)\end{equation}
where we have
\begin{equation}\label{theta-q}
q_0(t)=-\frac12\tilde\theta_x(0)+\frac12\int_0^t\sin\theta_0
\end{equation}
resulting from \eqref{check-k-sg}. 

\paragraph{Step 2:} then property P.3  holds for all $t$, namely
\begin{equation}
\Psi(k,0,t)=\left(\begin{array}{cc}1&0\\0&1\end{array}\right)\ .\end{equation}
The proof is obtained by following the
method of section \ref{sec:hilb-pro} where $q_0$ has simply to be
understood as $q_0(t)$, given from Dirichlet conditions by \eqref{theta-q}.

\paragraph{Step 3:} the limit of  $q(x,t)$ as $x\to0^+$ is computed out of the
expression \eqref{q(x)} by following the method of section \ref{sec:hilb-pro}
used to prove property P.8.

\paragraph{Step 4:} the limit of  $q=-\theta_x/2$ as $x\to\infty$ is computed
out of the expression \eqref{q(x)} again by following the proof of property
P.8.  Note that the essential property to be used is the fact that $\rho(k,t)$
is continuous and bounded for $k\in{\mathbb R}$ at any given $t$.  

\subsection{Summary of the method of solution of sine-Gordon}

The solution of sine-Gordon  with  Dirichlet conditions is achieved along the
following steps:

1 - The spectral transform at time zero, $\rho(k,0)$, is obtained form the
Dirichlet condition $\tilde\theta(x)$ by finding the vector solution 
$\phi_2(k,x,0)$ of the integral equations \eqref{int-phi+} in the upper half 
$k$-plane, namely by solving
\begin{align}
&\phi_{12}^+(k,x,0) =
-\int_{x}^{\infty}dy\ \tilde q(y) \phi_{22}^+(k,y,0)e^{-2ik(x-y )}
\notag\\
&\phi_{22}^+(k,x,0)=1+\int_{0}^{x}dy\  \tilde q(y) 
\phi_{12}^+(k,y,0)\ ,
\end{align}
where $\tilde q=-\tilde\theta_x/2$. Then $\rho(k,0)$ results from
\begin{equation}\label{rho-equiv}
\rho(k,0)=\left.\phi_{12}^+(k,x,0)\right|_{x=0^+}\ .\end{equation}

2 -  The spectral transform at time t, $\rho(k,t)$, is calculated in the upper 
half $k$-plane from $\rho(k,0)$ and Dirichlet condition $\theta_0(t)$
by  solving for $k\ne0$ the Riccati 
time-evolution 
\begin{equation}\label{riccati-sg}
\rho_t= \frac{i}{4k}\left[\rho^2\sin\theta_0-2\rho\cos\theta_0
-\sin\theta_0\right]\ ,\end{equation}
and for $k=0$
\begin{equation}
\rho(0,t)=\frac{\cos\theta_0-1}{\sin{\theta_0}}\ .
\end{equation}
The $N$ poles $k_n$ in the upper half-plane and related residues $\rho_n$ of
$\rho(k,t)$ must be identified for each value of the parameter $t$.

3 - The integral system \eqref{cauchy-psi-1}\eqref{cauchy-psi-2}
which, within the reduction \eqref{red-spectr-sg}, simplifies to
\begin{align}
\psi_{11}^+(k,x,t)=&1-\frac1{2i\pi}
\int_{-\infty-i0}^{+\infty-i0}\frac{d\lambda}{\lambda-k}\ 
\rho(-\lambda,t)e^{2i\lambda x}\psi_{12}^-(\lambda,x,t)
\notag\\
&-\sum_{n=1}^{N(t)}\frac{\rho_n(t)}{k_n(t)+k}
\psi_{12}^-(-k_n(t),x,t)e^{-2ik_n(t)x}\ ,\end{align}
\begin{align}
\psi_{12}^-(k,x,t)=&\frac1{2i\pi}
\int_{-\infty+i0}^{+\infty+i0}\frac{d\lambda}{\lambda-k}\ 
\rho(\lambda,t)e^{-2i\lambda x}\psi_{11}^+(\lambda,x,t) \notag\\
&-\sum_{n=1}^{N(t)}\frac{\rho_n(t)}{k_n(t)-k}\psi_{11}^+(k_n,x,t)e^{-2ik_n(t)x}
\ ,\end{align}
is solved to get $\psi_{11}^+(k,x,t)$.  This system of equation holds for any
fixed value of the {\em ``parameters''} $x$ and $t$.

4 - Finally the solution $\theta(x,t)$ is obtained under the form
$\theta_x(x,t)=-2q(x,t)$ where $q(x,t)$ results from $\psi_{11}^+$ and
$\rho$ as
\begin{align}
q(x,t)=&-\frac1\pi\int_{-\infty+i0}^{+\infty+i0}d\lambda\ 
\rho(\lambda,t)\psi_{11}^+(\lambda,x,t)e^{-2i\lambda x}\notag\\
&+2i\sum_{n=1}^{N(t)}\rho_n(t)\psi_{11}^+(k_n(t),x,t)e^{-2ik_n(t)x}\ .
\end{align}

\subsection{An explicit example}

To illustrate the above method we consider the explicitely solvable case
of a piecewise constant boundary datum $\theta_0(t)$, namely
\begin{align}
 t\in[0,t_1]\ :&\ \theta_0(t)=0\ ,\notag\\
 t\in[t_j,t_{j+1}]\ :&\ \theta_0(t)=\varphi_j\ ,\\
 t\in[t_n,\infty]\ :&\ \theta_0(t)=0\ ,\notag
\end{align}
compatible with a vanishing initial condition
\begin{equation}\theta(x,0)=\tilde\theta(x)=0\ .\end{equation}
Note that such a boundary datum represents an arbitrary localized function 
$\theta_0(t)$ discretized with $n$-step time grid.

We adopt the following notations
\begin{align}
&\rho_j(t)=\rho(k,t)\quad {\rm for}\quad t\in[t_j,t_{j+1}]\ ,\notag\\
&\rho_j=\rho_j(t_{j+1})\equiv\rho(k,t_{j+1})\ .
\end{align}
Then the solution of the evolution \eqref{evol-rho-sg} can be written
for $t\in[t_j,t_{j+1}]$
\begin{equation}\label{expsol}
\rho_j(t)=-\frac{b_j(\rho_{j-1}+a_j)e^{i(t-t_j)/2k}-a_j(\rho_{j-1}+b_j)}
{(\rho_{j-1}+a_j)e^{i(t-t_j)/2k}-(\rho_{j-1}+b_j)}\ ,
\end{equation}
with the following definitions
\begin{equation}
a_j=-\frac{1+\cos\varphi_j}{\sin\varphi_j}\ ,\quad
b_j=\frac{1-\cos\varphi_j}{\sin\varphi_j}\ .
\end{equation}
Note that the limit $\varphi_j\to0$ out of \eqref{expsol} does produces the
solution to evolution \eqref{evol-rho-sg} where $\theta_0=\varphi_j$ is set
to zero.

The large-k behaviour of solution (\ref{expsol}) is obtain by induction and
 reads
\begin{equation}
\rho_j(t)=-\frac{i}{4k}\left[(t-t_j)\sin\varphi_j+\sum_{\ell=1}^j
(t_\ell-t_{\ell-1})\sin\varphi_{\ell-1}\right]\ +{\cal O}(\frac{1}{k^2})\ .
\end{equation}
Hence we have
\begin{equation}
\partial_t\rho_j(t)=-\frac{i}{4k}\sin\varphi_j+{\cal O}(\frac{1}{k^2})
\end{equation}
compatible with (\ref{check-k-sg}) and solution 
(\ref{expsol}) has the expected large-k behaviour.

At the last step, that is for $\varphi_n=0$, \eqref{expsol} can be writen
\begin{equation}
\rho_n(t)=-\frac{(b_n\rho_{n-1}-1)e^{i(t-t_n)/2k}-a_n\rho_{n-1}+1}
{(\rho_{n-1}+a_n)e^{i(t-t_n)/2k}-(\rho_{n-1}+b_n)}\ ,
\end{equation}
which eventually simplifies to
\begin{equation}\rho_n(t)=\rho_{n-1}e^{i(t-t_n)/2k}\ .\end{equation}
As a consequence, the spectral transform evaluated {\em after} the total effect
of the boundary datum $\theta_0(t)$, i.e. for $t>t_n$, is completely determined
by $\rho_{n-1}=\rho(k,t_n)$. This function of $k$ is itself obtained as
the following {\em mapping}
\begin{align}\label{mapping}
&\rho_0=0\ ,\notag\\
&\rho_j=-\frac{b_j(\rho_{j-1}+a_j)e^{i(t_{j+1}-t_j)/2k}-a_j(\rho_{j-1}+b_j)}
{(\rho_{j-1}+a_j)e^{i(t_{j+1}-t_j)/2k}-(\rho_{j-1}+b_j)}\ ,
\ j=1\cdots n-1\ .
\end{align}
This is an explicit expression that allows to get insight in the effect of
a generic input datum $\theta_0(t)$ on a vanishing background under
suitable (arbitrary) discretization. We report the study of the above mapping
(in the complex $k$-plane) to forthcoming work and simply mention here that
the main point concerns the study of the creation (anihilation) of poles
of $\rho(k,t)$ in the upper half $k$-plane.

Remembering the reality constraint \eqref{real-red-sg-poles}, poles of
$\rho(k)$ come by pair, and a pole $k=k_n(t)$ of $\rho_{j}(t)$  is given by 
\begin{equation}\label{polerho}
e^{i\frac{t-t_j}{2k_n}}=\frac{\rho_{j-1}+b_j}{\rho_{j-1}+a_j}\ ,
\end{equation}
where $\rho_{j-1}$ is itself a fuction of $k_n$.  

This equation for the pole $k_n$ is in general highly complicated but if
$\rho_{j-1}=0$ (as for the initial stage $t=t_1$), the above relation becomes
\begin{equation}\label{initial}
e^{i\frac{t-t_1}{2k_n}}=-\frac{1-\cos\varphi_1}{1+\cos\varphi_1}\ .
\end{equation}
Defining 
\begin{equation}\gamma=\ln\frac{1-\cos\varphi_1}{1+\cos\varphi_1}\ ,
\end{equation}
the solution of \eqref{initial} can be written
\begin{equation}\label{k-n}
k_n(t)=\frac12\left[(2n+1)\pi+i\gamma\right]
\frac{(t-t_1)}{(2n+1)^2\pi^2+\gamma^2}\ ,
\end{equation}
where $n$ assumes values in $\mathbb Z$. The above expression is identical to
the one obtained in \cite{fokas} (see formula 5.52) for $t-t_1=T$. A
property of our approach is to explictely display the {\em time dynamics} of  
the spectral transform (such as poles generation and motion).
\vskip10pt
\begin{figure}[ht]\begin{center}
\epsfig{file=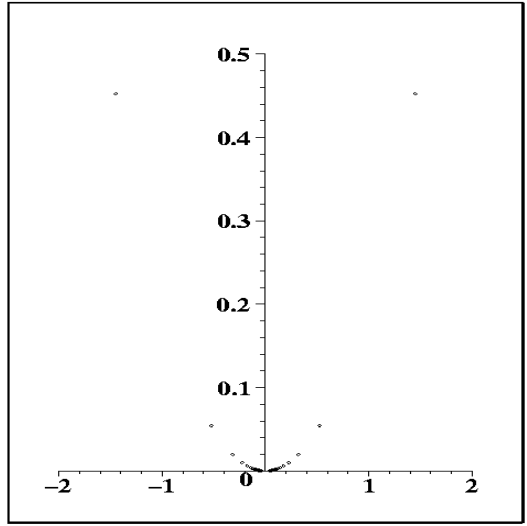,height=7cm,width=7cm}
\caption{Positions of the poles of $\rho(k,t)$ for $n\in[-20,19]$, $t-t_1=10$
and $\varphi_1=0.65\pi$}\label{poles} 
\end{center}\end{figure}
Among the above set of poles, only those
with positive imaginary part contribute to the soliton spectrum. It is then
remarkable that 
\begin{align}
&\varphi_1\in[0,\frac\pi2[\ \Rightarrow\ \gamma<0\ \Rightarrow
\ {\rm Im}(k_n)<0 \ ,\\
&\varphi_1\in]\frac\pi2,\pi]\ \Rightarrow\ \gamma>0\ \Rightarrow
\ {\rm Im}(k_n)>0\ ,\end{align}
which shows that no poles are generated by a consant boundary datum
$\theta_0(t)=\varphi_1$ smaller than $\pi/2$, while an infinity of them are
generated for $\varphi_1>\pi/2$. An example of poles locations is shown on the
figure where we see that the point $k=0$ is an accumulation point for the
infinity of poles.  Indeed when $n\to\infty$ we have from \eqref{k-n}
$k_n\to0$. The case $\varphi_1=\pi/2$ is singular as $\rho(k,t)$ would then
possess poles on the real axis, for which the theory fails.

Note that the discretization of a smooth localized boundary datum $\theta_0(t)$
produces {\em small} $\varphi_j$'s. Hence the soliton generation would then
result from the contributions in equation \eqref{polerho} of the
$\rho_{j-\ell}$. In the case of a single constant boundary datun of amplitude
greter than $\pi/2$, the assignation of solitons to the presence of an infinity
of poles is not clear (usually an infinity of poles is related to a
self-similar structure).

\section*{Aknowledgements.} 
It is a pleasure to aknowledge enlightening discussions with M. Boiti, 
E.V. Doktorov, A.V. Mikhailov, B. Pelloni, F. Pempinelli, P.C. Sabatier and 
V.S. Shchesnovich. This work has been partially suported by the contract
INTAS 99-1782.

\end{document}